\documentclass[preprint,pra,amssymb,amsmath,showpacs,floatfix]{revtex4}

\usepackage{graphicx}
\usepackage{amssymb}
\usepackage{amsmath}

\newcommand{\be}{\begin{equation}}
\newcommand{\ee}{\end{equation}}
\newcommand{\bea}{\begin{eqnarray}}
\newcommand{\eea}{\end{eqnarray}}

\begin{document}

\title{ Why are Probabilistic Laws Governing Quantum Mechanics 
and Neurobiology? }

\author{
Helmut Kr\"{o}ger,\footnote{E-mail: hkroger@phy.ulaval.ca} 
}

\affiliation{
{\small\sl D\'epartement de Physique, Universit\'e Laval, Qu\'ebec, 
Qu\'ebec G1K 7P4, Canada} 
}

\begin{abstract}
We address the question: Why are dynamical laws 
governing in quantum mechanics and in neuroscience of probabilistic nature instead of being deterministic? 
We discuss some ideas showing that the probabilistic option offers 
advantages over the deterministic one. 
\end{abstract}

\maketitle

\setcounter{page}{0}
\newpage

\section{Overview}
\label{sec:Over}
\noindent Neuroscience is part of biology. In biology, Darwin's theory of evolution of species - which corresponds to the standard model in elementary particle physics -
is based on two principles: The system undergoes small changes - encoded in the genes - which are random. Then there is competition (fighting, survival of the fittest) which means selection and which eventually may lead to the emergence of new species. 
As time evolves, biological evolution has generated forms of life starting from 
sinlge cells, amoebia and eventually producing large animals (dinosaurs, whales). To use a modern term, complexity has increased. 
One may say that the random change is the motor which drives evolution, and 
selection takes care that complex forms of life emerge.

If we look at the evolution of the universe, the creation of elementary particles, the formation of nuclei (nucleosynthesis), the formation of atoms, and the formation of macromolecules like proteins, we notice that there is increasing complexity observable at different levels of length or energy (physicists often talk about scales of length or scales of energy, like Planck length, $\Lambda_{QCD}$, binding energy of proton, binding energy of $^{4}He$, binding energy of atoms, of molecules etc.).
Like in Darwin's theory, the concept of randomness is present in the 
evolution of the universe. For example, in the big-bang standard model of cosmology, the universe starts out from extremely high temperature where the laws of thermodynamics are believed to be valid. Thus the universe is described in terms of statistical field theory, where fluctuations occur in a random manner. During the early expansion phase, the inflationay model assumes that some tunneling transition has occured, which is of probabilistic nature.

The concept of probability and chance can be viewed as a motor
driving the evolution of the living organisms but also the evolution of our universe. The essential difference between both is the mechanism of emergence of complexity. The problem of emergence of complexity in the universe is related to the well known fine-tuning problem in nature:
It means that there are some constants of nature the values of which necessarily have to lie in quite a small window, otherwise the complex universe as we see it today would not exist. Examples are windows for quark masses, windows for the cosmological constant, windows for the dipole moment of water and windows for a lot of other quantities. 
The question why such constants of nature take on values in such windows is an  unresolved puzzle. Recently, this puzzle has become entangled with the observation that some constants of nature - like the fine structure constant $\alpha$ - have varied in time during the evolution of the universe \cite{kn:Constants}. The subject of this article is the question: Why does nature use the concepts of probability and chance in contrast to the alternative of determinism?

\section{Introduction}
\label{sec:Introduction}
\noindent We start out by asking the question: Are the basic laws governing in neurobiology of probabilistic or of deterministic nature? What if we ask the same question in quantum physics? The answer is well known for quantum mechanics: The laws are of probabilistic nature. In neurobiology one has not such a clear cut answer. But there is much evidence in favor of stochastic behavior. 
For example, there is irregular spiking behavior of cortical neurons in vivo.
There is noise in synaptic transmission. 
Noise plays a role in the working of ion channels in the neural membrane.
The reader might wonder why we are going to treat such different 
topics as quantum mechanics and neuroscience on the same footing, 
although it is generally believed that quantum 
mechanics is not needed to understand the working of the brain or of neuroscience in general. 
(An opposite view has been taken by Penrose \cite{kn:Penrose}). 
The reasons are, first, that the typical length scales of Q.M. are 
similar to those playing a role in the working of a 
neuron. Secondly, the answers we are offering have something in 
common for Q.M. and neuroscience.

In the following we will focus on the question: 
Why are those laws of stochastic and not of deterministic nature? 
The reader might wonder in the first place: What makes us ask this question?
Like in a detective story, asking the motives of a suspect often 
helps to trace the history of the crime and solve 
the murder puzzle. Likewise we hope that those questions will help to better understand neuroscience. Of course we can not give an answer to the main question. However, we will propose a tentative answer in the sense that it is favorable for nature to use the probabilistic option. 
We will present arguments in support of this thesis.

\section{Tentative answer}
\label{sec:Tentative}
\noindent The tentative answer which we propose for both Q.M. and neuroscience is: 
Nature has chosen the probabilistic option, because it offers the following advantages: \\
\\
Quantum mechanics: \\
(1) Architectural simplicity and efficiency. \\
(2) Algorithmical simplicity. \\
(3) Cost efficiency. \\
(4) Repair efficiency and robustness. \\
(5) Infinite lifetime (in principle). \\
Neuroscience: \\
(1) Efficient (and may be the only) strategy 
for adaptive learning. \\ 
(2) Random connections are part of the small-world and scale-free network achitecture, which both were shown to be advantageous in nature.
\\
\\
\noindent Before entering into discussion, 
let us make a brief historic note. According to scientific evidence Q.M. 
and also neurobiology at the level of individual cells  
is ruled by stochastic laws. On the other hand, we know that the 
macroscopic world is described by classical physics, which is of 
deterministic nature. Historically, in the era of 
renaissance and rationalism, scientists believed that the whole 
universe could be described in terms of deterministic laws. 
With the advent of quantum mechanics, and the concept of probability 
involved, many people and in particular philosophers had great 
difficulties to accept that. Even Einstein believed in deterministic 
laws underlying quantum mechanics: "...Gott w\"urfelt nicht..." (God does not role the dice).

This development of view and thought is quite natural, however. 
In the evolution of humans over the last 3 million years, man has 
investigated nature at the macroscopic scale, i.e. at the scale 
of resolution of the bare eye and length scales between resolution of 
touching sense to walking distance. Nature at the scale of atoms or 
neurons was not accessible to man until, say the last century. 
So mankind described nature in terms of classical physics.
The interesting point is: When it became clear that 
probabilistic laws are at work at the length scale of Q.M., 
how did that change scientific thought?
One direction was (Einstein): Shouldn't quantum mechanical laws be explained 
in terms of underlying deterministic laws? This direction 
corresponds to explain nature at the microscopic level by laws
describing nature at the macroscopic level where nature is much more complex.  
Sofar there is no scientific 
evidence that this is possible.

Scientists also thought about the other direction: 
Can the deterministic laws of macroscopic physics be explained in 
terms of probabilistic laws of microscopic physics?
This has been much more fruitful, as the success of statistical 
mechanics has demonstrated. 
Most scientists believe that the answer is yes!
This direction corresponds to describe nature at the complex macroscopic level
by reducing it to the laws of nature at the less complex microscopic level.

\section{Example of traffic: deterministic versus stochastic dynamics}
\label{sec:Traffic}
\begin{figure}[tph]
\vspace{9pt}
\begin{center}
\includegraphics[scale=0.6,angle=0]{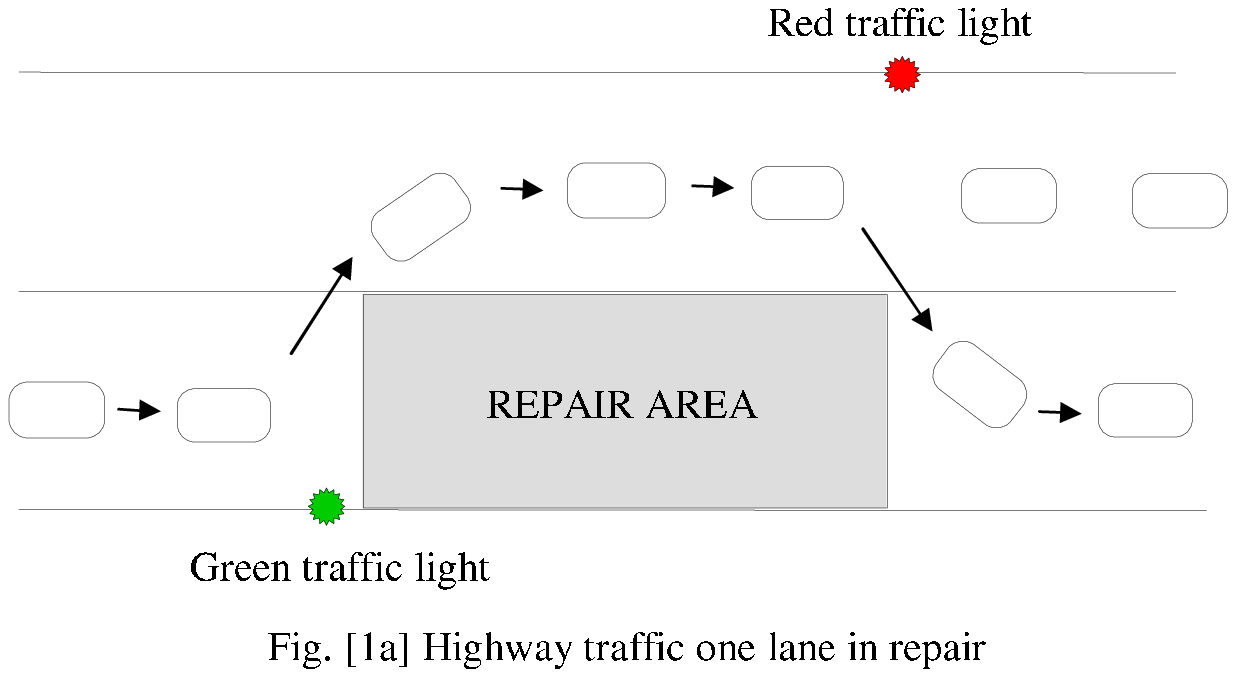}
\includegraphics[scale=0.6,angle=0]{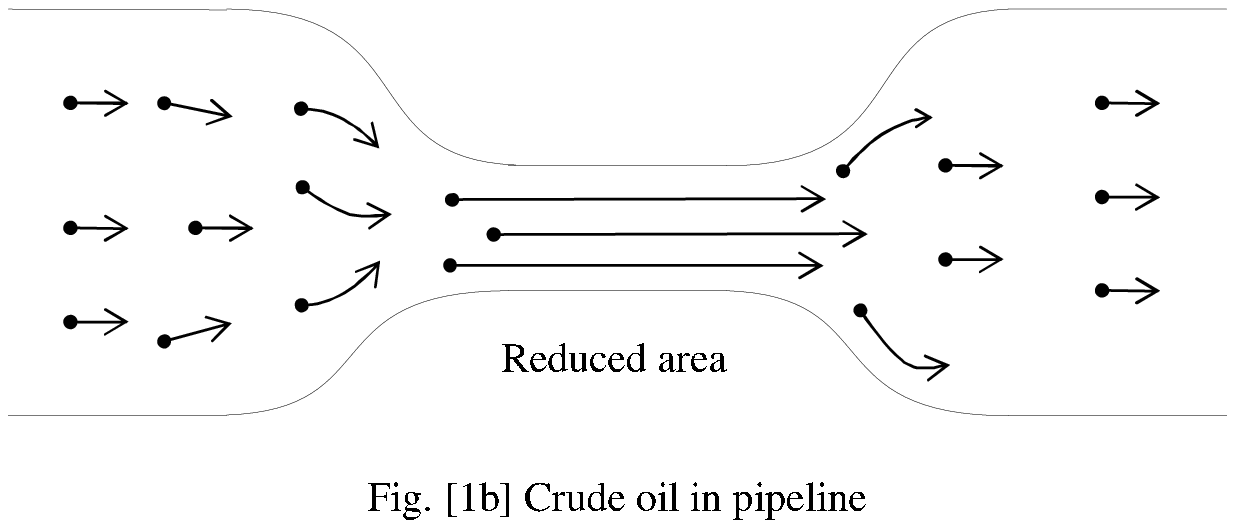}
\end{center}
\caption{Flow of car traffic on highway around repair area.
Schema of flow of car traffic on highway (a) 
compared to flow of oil particles in pipeline (b).}
\label{fig:Traffic}
\end{figure}

\noindent Let us start by considering the following example 
well known from daily life, i.e., from the macroscopic world.
Fig.[\ref{fig:Traffic}a] shows automobile traffic on a two lane highway, with 
traffic passing an area of road work, with one lane being closed. At 
both ends of the repair area, traffic lights control the motion 
of traffic in both directions. At a given time traffic moves only 
in one direction. In this case, traffic is governed by a 
deterministic law (one can exactly predict, when the traffic light will change).

In Fig.[\ref{fig:Traffic}b] we consider crude oil travelling 
in a pipeline, passing an area of reduced diameter. Oil molecules behave 
quite different from automobiles. They carry out Brownian motion with an 
underlying constant motion along the pipeline. This is a stochastic process. 
For an ideal (incompressible) liquid holds Bernoulli's equation
$p + \frac{1}{2} \rho u^{2} = \mbox{const}$, where $p$ is the pressure, $\rho$ is the density and $u$ the velocity of the liquid. Moreover, there is a simple equation of continuity
\begin{equation}
F_{1} u_{1} = F_{2} u_{2} .
\end{equation}
It means, when the liquid passes areas where the cross section of the tube is $F_{1}$ and $F_{2}$, respectively, then the velocity accomodates such that 
product of velocity and cross section is constant.
Consequently, the oil before and after the area of reduced diameter travels 
with the same speed. When the molecules reach the reduced
section of the tube, particles will collide, but they begin to move faster.
Comparing scenario (a) with (b), we find that in scenario (a) the traffic 
moves slower compared to when the repair area is absent. 
In scenario (b), the oil moves faster in the narrow zone and after having passed the narrow zone it moves with the same velocity as before that zone.

This is a first example showing that a system with underlying 
stochastic dynamics wins over a similar system with deterministic dynamics. 
This example demonstrates point (1): The traffic light supervision system 
is much more complicated than the pipeline, where a supervision 
mechanism  is absent. Secondly, it is relatively slower.
Of course scenario (b) holds only in a certain window of parameters. It 
depends on viscosity, pressure, velocity, Reynolds number etc. Under 
extreme conditions there is a transition to turbulence and the flow becomes jammed.

This example hints to some understanding why stochastic behavior wins over deterministic behavior. The oil molecules (although
being large molecules) are objects much simpler structure than cars.
In particular, when undergoing collisions, the oil molecules come out undamaged,
while cars (being complicated complex objects) would undergo heavy damage. So we formulate this observation: 
Microscopic objects of low-level complexity show stochastic behavior, while macroscopic objects of high-level complexity follow deterministic behavior.
But aren't there counter examples to this claim? Let us consider the motion of ants. They apparently move very much like a random walker. And aren't ants beings of high complexity?
Yes, they certainly are. However, their motion is not purely random. For instance when several ants want to enter into the ant hill at the same time, they do it in an orderly manner (more like the cars than the oil molecules) \cite{kn:Ants}.

\section{Probabilistic nature of quantum mechanics}
\label{sec:Quantum}
\subsection{Single particle events in scattering}
\label{sec:SingleEvents}
\noindent In Q.M. the occurence of single events is purely random. For example, consider the emission of a photon from a light source. The time instant of emission can not be predicted. Why is it a random event?
One can argue that, on the contrary, if it were not random or probabilistic, but deterministic, it would lead to contradictions in Q.M. As example let us consider Young's double slit experiment showing an interference effect of electrons.
There is a source emitting electrons (like such existing in a television tube.
The emission of electrons from a metal surface can be achieved by heating the metal). The electrons impinge on the screen A which has two holes.
Some electrons get stuck. Some pass through the holes.
Those which have passed continue and move on towards the second screen B. 
Some of those end up in the detector and are counted. 
The whole set up can be considered as a scattering experiment, where the beam consists of electrons and the target is represented by the screen A with 
two holes.
In doing such experiment and counting the number of electrons (intensity) 
in the detector as a function of position $x$ of the detector, one observes a curve with a maximum at the center $x=0$. However there are several lower
side maxima. Also there are several minima, corresponmding to intensity zero.

This is an interference pattern. Why is it called an interference pattern?
Because it gives the same pattern, which would have been obtained in studying the behavior of water waves.
Suppose one creates spherical water waves by periodically exerting pressure at the same place on the surface of water. These water waves propagate and arrive at the screen A (with holes positioned such that half of their opening is above water). Each hole creates a new sperical wave, which both propagate. Those interact with each other, creating a wiggly surface of minima and maxima, which can be observed in the detector.

So it turns out that electrons are behaving just like water waves.
That is why quantum mechanics has historically been dubbed as wave mechanics.
However, there is one essential difference: Water waves create this interference pattern only when secondary spherical waves are created at the holes 
in coincidence. That means the time instant when the spherical wave of the source arrives at hole number one is not very different from the instant when it arrives at hole number two.  
This is not necessarily so in quantum mechanics.
First, when doing the quantum mechanical experiment with electrons, one 
can verify experimentally, that it is actually one electron at a time which passes a hole. And this electron passes either the upper hole or the lower hole
at a time. Now comes the surprise. One can tune the electron source such that it emits an electron at a time with very long silent intervals in between.
Nevertheless one observes that each electron contributes to build in the detector the interference pattern.
With what did the electron interfer?
If we would interpret the electron as an object obeying the laws of classical physics, 
there would be nothing the electron could interfer with, hence it 
hardly could generate an interference pattern. 
On the other hand, considering the electron obeying the laws of quantum mechanics, it is described by a wave function. The wave functioon can be decomposed in a part corresponding to the passage by hole one and a part corresponding to the passage of hole two. Then the interference pattern can be
obtained by interference of the two pieces of wave function. 
The wave function $\psi(x)$ has an interpretation as probability amplitude.
The probability itself is given by $P(x) = |\psi(x)|^{2}$ (see sect. \ref{sec:Concepts}). In conclusion, the scattering process of an electron from 
screen with two holes leads to into a conflict, when adopting the classical physics, i.e. deterministic, point of view to explain the observed interference pattern.

\subsection{Probabilistic decay of radio nuclei}
\label{sec:ProbDecay}
\noindent Another similar example for the probabilistic nature of 
quantum mechanics is the decay of radio nuclei. E.g., let us consider an Uranium atom $^{235}U$. It decays via certain pathways into a number of decay products, each decay channel being associated with a certain average life time.
The question is: Why does it decay? What is the underlying mechanism? 
Of course we do have a quantum mechanical explanation for the decay. 
But we do not know of any deterministic law which tells us for a given 
particular $^{235}U$ atom, when exactly it will decay. There is general consensus that such law does not exist.

\begin{figure}[tph]
\vspace{9pt}
\begin{center}
\includegraphics[scale=0.6,angle=0]{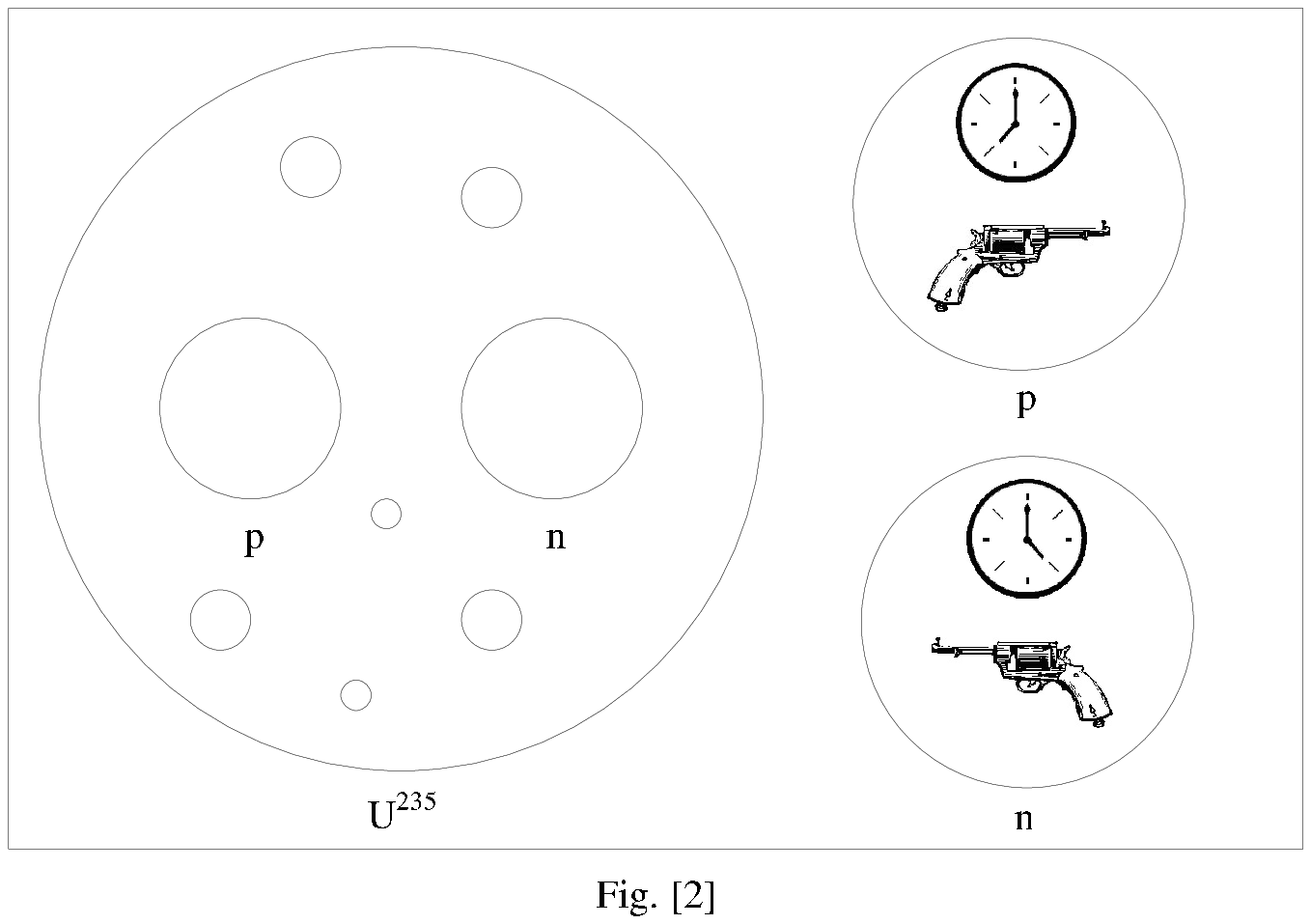}
\end{center}
\caption{
Scenario of deterministic decay of $^{235}U$ nucleus.
}
\label{fig:UDecay}
\end{figure}

Why is there no such deterministic law? 
Let us put ourselves in position of the creator of an $^{235}U$ atom: 
Can we conceive a deterministic mechanism for the $^{235}U$ atom to have the decay 
properties observed in nature? 
How could that look like? The Uranium atom $^{235}U$
(see Fig.[\ref{fig:UDecay}]) is built from protons and neutrons (the total number of which is 235). 
Let us consider the following scenario: Suppose in the interior of each proton and neutron there is a clock and a loaded gun (similar to Schr\"odinger's cat paradoxon). At a certain time, the moment of decay, the alarm rings. The alarm is connected with the gun. The gun fires a bullet onto some partner nucleon, giving it some momentum, which is sufficient to overcome the potential barrier and the atom decays. 

Of course every child knows that this is fiction! But we ask: 
Why is this not realized in nature?
The answer is: It would be way too complicated!
The alarm clock and the gun are both objects of a much higher level 
of complexity then the proton and the neutron.
So we ask: Can we do better and conceive a deterministic 
mechanism which is more simple 
(comparable to the level of complexity of the proton and the neutron)?
Note that the most complicated object involved is the clock: Atomic 
clocks exist, which measures the 
oscillations of a $Cs$-atom. It requires a very complicated 
experimental set up in the laboratory.
The alarm clock and the gun are objects of a much higher level of structure and complexity than the proton and the neutron.
Note, that the $^{235}U$ atom or similar atoms may have quite a long life time 
(compared to intrinsic energy or time scales). If there would be an 
internal clock, it would need a memory device to store (count) the 
time units. It should be able to memorize time for quite a long time! 
Such a device would have a complex structure.

Can we construct a clock which 
is precise but no more complicated than an atom? In the $Cs$ atomic 
clock, the $Cs$ atom is not complicated, but the measuring apparatus 
is! Measurement means interaction. So we need an interaction. 
Let us consider the nucleon spin. The spin of a nucleon is aligned with the 
magnetic field if an external strong magnetic field is applied. 
The pointer of the clock corresponds to the direction of the precessing spin.
But again measuring the direction of spin at a fixed time requires 
an apparatus much more complicated than a single atom!

\begin{figure}[tph]
\vspace{9pt}
\begin{center}
\includegraphics[scale=0.6,angle=0]{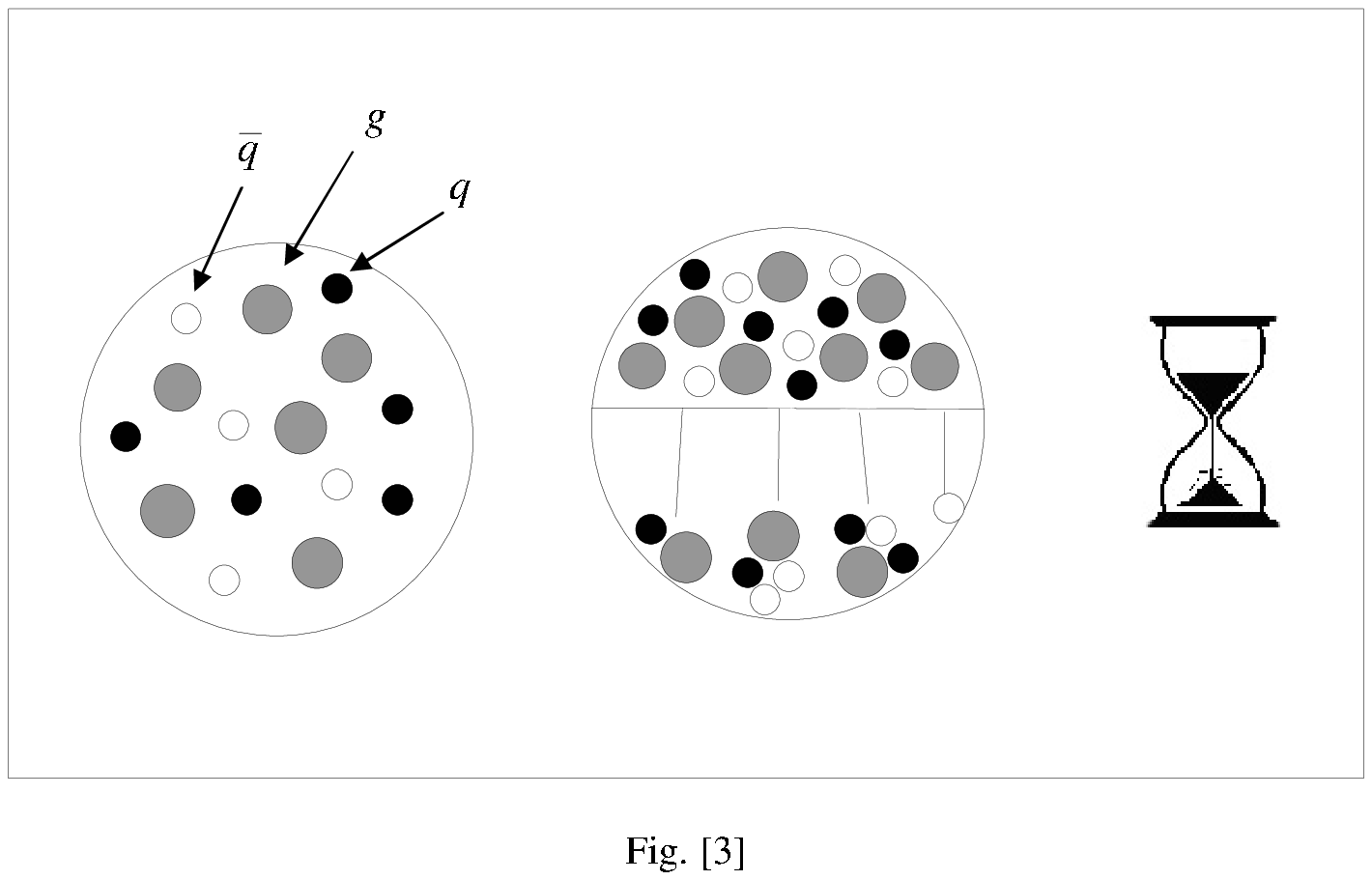}
\end{center}
\caption{
Schema of sand-clock of hadron, where quarks and gluons represent 
the "grains of sand". 
}
\label{fig:SandClock}
\end{figure}

We can imagine another possibility: 
It is well known that the nucleon has 
a substructure: Quarks and gluons. Imagine the "clock" to work like a 
sand-clock, but instead of sand grains utilizing quarks and gluons (see Fig.[\ref{fig:SandClock}]). However, the motion of "sand grains" is more like a stochastic than a deterministic process. 
One could conceive many other possibilities of clocks and 
for the decay process. We claim that either the clock is much more 
complicated than the atom or the clock is based on a stochastic process.

\begin{figure}[tph]
\vspace{9pt}
\begin{center}
\includegraphics[scale=0.6,angle=0]{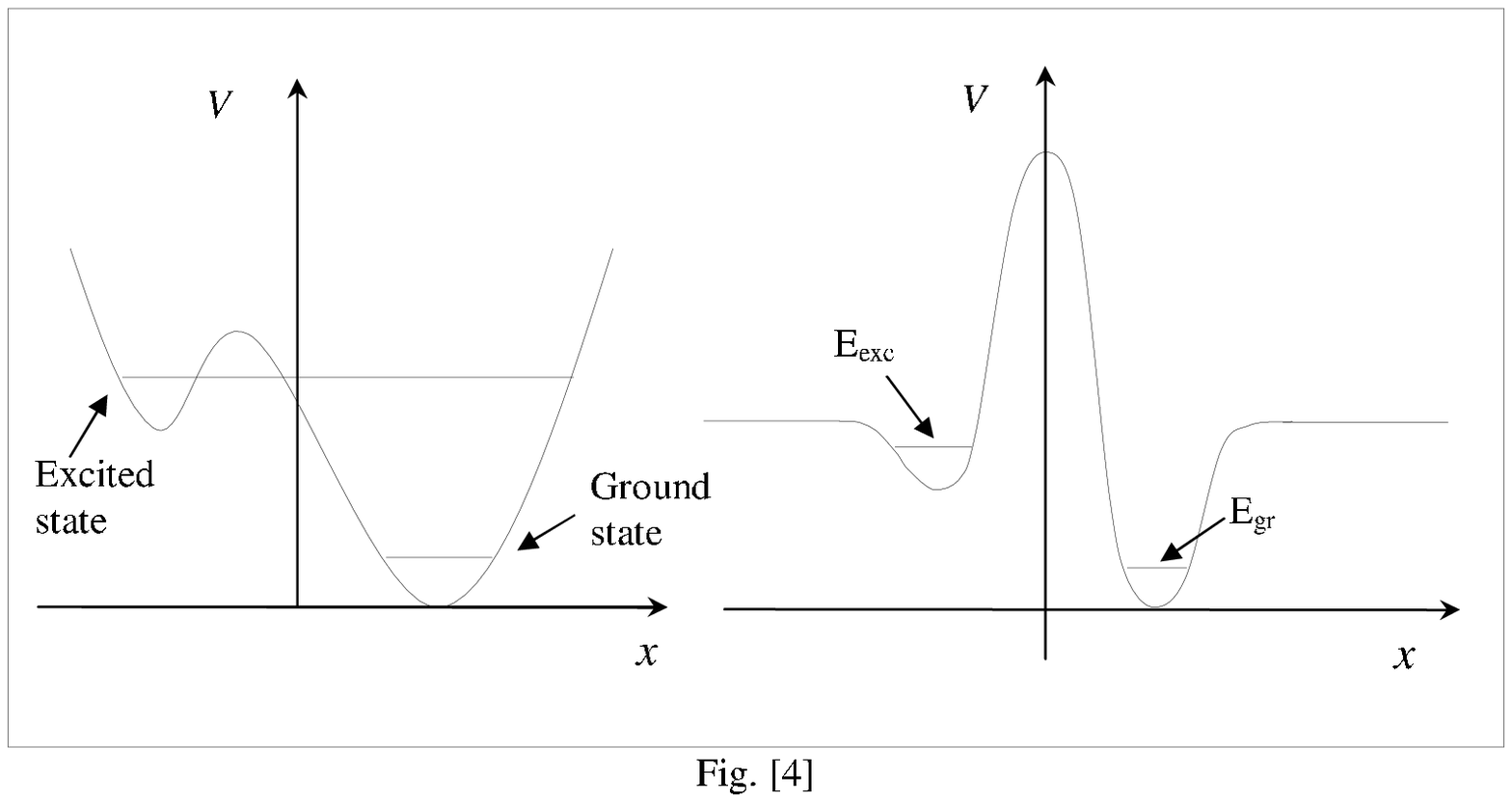}
\end{center}
\caption{
Potential model for Uranium decay as a transition from an 
unstable excited state to a stable ground state. 
}
\label{fig:UPotModel}
\end{figure}

Let us see, on the contrary, how the decay can be understood in a 
quantum mechanical, i.e. probabilistic, model. Consider Fig.[\ref{fig:UPotModel}]. 
There is a ground state and an excited state, represented by absolute 
and relative minima of the potential. The excited state is unstable. 
When the potential barrier is sufficiently high, the excited state 
is long living. In the limit of an infinite barrier height, the excited 
state becomes a stable state. The probabilistic laws of quantum 
mechanics allow for a transition from the excited state to the ground 
state (similar to tunneling). Note, that there is no additional mechanism 
necessary! The effect comes from the probabilistic nature of the 
wave function. Remark: A similar situation prevails, when considering 
a high lying excited state of, say a $H$ atom (Rydberg states) with an 
excitation transition induced by photo absorption (laser) and 
de-excitation by photo-emission. The transition occurs due to the 
interaction with an electromagnetic field. 
Lesson: Q.M. predicts the transition probability and the decay 
rates. But it {\em does not} provide a deterministic statement 
of the precise decay time of a particular atom. The trade-off is: 
Q.M. probabilistic laws are much simpler then a deterministic 
law would have been!

\subsection{What if laws of physics at the atomic scale were deterministic?}
\label{sec:World}
In the real world, we believe that nature at atomic 
scales is described by quantum mechanics. There is a wave function 
$\psi(\vec{x},t)$ which is a complex number (for any given value of $\vec{x}$ and $t$). It has a probabilistic interpretation.
\begin{equation}
P = |\psi(\vec{x},t)|^{2} \Delta V
\end{equation}
gives the probability to find a particle, obeying the Schr\"odinger 
equation with a wave function $\psi$, in a space volume $\Delta V$ 
around the position $\vec{x}$ at time $t$.

Since the advent of quantum mechanics, we can explain 
the discrete levels of bound states in atoms, which classical physics, 
based on deterministic laws, failed to explain. 
This was the first great success of quantum theory!
We ask: What if an atom would be a classical object, governed by 
deterministic laws? Firstly, one can look from the point of view of 
classical chaos. It has been known since Poincar\'e 
that a classical 3-body system may be a chaotic system.   
Thus our solar system may not be stable. 
Numerical solutions of orbits by Sussman and Wisdom \cite{kn:Sussman88}, Laskar \cite{kn:Laskar89} and recent analytic calculations by Murray and Holman \cite{kn:Murray99} show that the jovian planets Jupiter, Saturn, Uranus and Neptune are chaotic, with an estimated Lyapunov life time of $10^{7}$ years. 
If we consider a 3-body system at the atomic scale, 
an example would be the deuterium atom (heavy water) consisting 
of a proton, a neutron and an electron.
Under the assumption that such a system would be ruled by classical laws, 
it would likely be chaotic (it is not quite the same as the celestical 
3-body system, because the gravitational force is $F \propto 1/r^{2}$, 
which is if the same type as the Coulomb force, but the strong force 
between proton and neutron is not of this type). 
More complicated objects like large molecules with a complicated binding 
mechanism are quantum mechanically stable. If such a system would be 
governed by deterministic laws, it would quite likely lead to collisions 
and strong chaotic behavior, leading eventually to decay, which means 
a short life time of such objects. 

Secondly, one can look from the point of view of classical 
electrodynamics. In classical physics, the hydrogen atom is a 2-body system 
composed of a heavy proton and a light electron, both carrying electric 
charge, and the electron orbits around the proton. The orbiting electron 
follows a curved trajectory, which means the particle undergoes acceleration. 
Classical electrodynamics predicts that this causes radiation, which goes 
with a loss of kinetic energy. As a consequence, the atom would not be 
stable and collapse rapidly. This effect is much stronger than the chaotic 
dynamics effect. The electrodynamical radiation loss and its prediction of 
unstable atoms was an import incentive for the invention of quantum mechanics. 

As a result, if an atom would obey the laws of classical physics, 
it would be an unstable object. No macromolecules (proteins) would exist 
in nature. DNA would not exist, and hence no organic life.

\subsection{Existence of stable atoms and probability}
\label{sec:StableAtoms}
\noindent We want to show that the existence of stable atoms can be traced back to the concept of probability in quantum mechanics.
A stable atom means that there are discrete energy levels 
$E_{0} < E_{1} < E_{2} ...$, with gaps $\Delta E_{i} = E_{i} - E_{i-1} > 0$. Once the atom occupies one of those states, it can stay in this state forever (if there are no interactions with any other atoms or any electromagnetic field). In classical mechanics as well as in Q.M., the Hamiltonian is given by kinetic energy plus potential energy,
\begin{equation}
H = T + V .
\end{equation}
We claim that a discrete spectrum is possible only when $T$ and $V$ do not commute. Mathematically, this means
\begin{equation}
\label{TVnoncommute}
T V - V T \equiv [T,V] \neq 0 .
\end{equation}
Physically, this means that kinetic energy and potential energy can not be measured simultaneously.
\\

\noindent {\bf Theorem.} $H = T + V$ has a discrete spectrum only if $[T,V] \neq 0$. \\
{\bf Proof} (heuristic). Let us assume that $T$ and $V$ commute,
\begin{equation}
[T,V] = 0 .
\end{equation}
There is a mathematical theorem, stating that if two operators commute, then they can both be diagonalized in a common basis. The kinetic energy operator $T = \frac{\vec{P}^{2}}{2m}$ is diagonalized in a momentum basis 
\begin{equation}
T | \vec{p} > = \frac{\vec{p}^{2}}{2m} | \vec{p} > .
\end{equation}
Hence also $V$ must be diagonal in this basis,
\begin{equation}
V | \vec{p} > = {\it v}(\vec{p}) | \vec{p} > .
\end{equation}
The Schr\"odinger equation implies
\begin{equation}
\left( \frac{\vec{p}^{2}}{2m} + {\it v}(\vec{p}) \right) | \vec{p} > =  E_{p} | \vec{p} > .
\end{equation}
Thus we find for the spectrum
\begin{equation} 
E_{p} = \frac{\vec{p}^{2}}{2m} + 
{\it v}(\vec{p}) .
\end{equation}
Making the reasonable assumption that the potential is at least a piece-wise continuous function,
one obtains that the spectrum $E_{p}$ may have a multi-band structure, but certainly is not compatible with the structure of individual discrete levels with finite gaps, which proves the theorem.

In a one-body system one has
\begin{equation}
T = \frac{\vec{p}^{2}}{2 m}, ~~~ V = V(\vec{x}) .
\end{equation}
Eq.(\ref{TVnoncommute}) is equivalent to 
\begin{equation}
[ \vec{X},\vec{P} ] \neq 0 .
\end{equation}
Actually, 
\begin{equation}
\label{XPcommutator}
[ X, P_{x} ] = i \hbar , ~~~ \mbox{same for y- and z-component}
\end{equation}
is the fundamental commutator relation between position and momentum in Q.M. The latter equation is closely related to Heisenberg's uncertainty relation 
\begin{equation}
\label{Heisenberg}
\Delta X ~ \Delta P_{x} \geq \frac{\hbar}{2}, ~~~ \mbox{same for y- and z-component}
\end{equation}
being fundamental property which has been experimentally observed. 
Actually, Eq.(\ref{Heisenberg}) can be derived from Eq.(\ref{XPcommutator}) \cite{kn:Messiah}.
Here $\Delta X$ is defined by 
\begin{eqnarray}
\Delta X &=& \sqrt{ <\psi| (X - \bar{X})^{2} | \psi> }
\nonumber \\
&=& \int d^{3}x (x - \bar{X})^{2} |\psi(\vec{x})|^{2} 
\nonumber \\
&=& \int d^{3}x (x - \bar{X})^{2} P(\vec{x}) ,
\end{eqnarray}
where $\bar{X} = <\psi|X|\psi>$ denotes the mean value of $X$ in the state $\psi$. $\Delta X$ is the root mean square deviation or variance of $X$ 
for a probability density distribution given by $P(x) = |<\vec{x}|\psi>|^{2}$. 
$\Delta P$ is defined correspondingly. As can be seen, both $\Delta X$
and $\Delta P$ depend on the particular wave function $\psi$. 
Heisenberg's uncertainty relation can be interpreted as an upper bound for any wave function $\psi$ on the product of variances of position and momentum.
This shows how the concept of non-commuting operators leads to the notion of mean and variance, both concept from probability theory.
This ends our discussion on the relation between stable states and probability in Q.M.

\section{Implication of probability on geometry}
\label{sec:Geometry}
Above we have discussed the relation between probability, Heisenberg's uncertainty relation, and the commutator between the position and momentum operator.
The occurrence of non-vanishing variance means that there are quantum fluctuations.
A well known example are zero-point fluctuations for the ground state energy,
which implies that classical and quantum ground state energy differ. Another example is the propagation of a quantum particle. Feynman's path integral describes the propagator as a sum of weights $exp[iS(x(t))/\hbar]$, where $S$ is the action and $x(t)$ is a path from the starting point to the end point of propagation. Infinitely many such paths contribute to the propagator. These paths can be viewed as fluctuations around the classical trajectory. The propagator can be viewed as the wave function corresponding to the specific initial condition that the particle is located initially at the starting point of propagation. This connects quantum fluctuations to probability of the wave function.

Now when we speak about geometry in quantum physics, we do not mean the standard coordinate system of position $\vec{x}$ and time $t$. But we mean to talk about geometry closely related to the dynamics of the system. This idea is basically Einsteins old idea to express forces of gravitation in terms of the geometry of space time. Can we do something similar in quantum physics? In an attempt to do this, one can consider paths occuring in the path integral of the propagator. We introduce a geometry in the space of paths by introducing a distance of paths. We can do this in the following way:
Say we have two paths, $x_{1}(t)$ and $x_{2}(t)$. There are two corresponding weight factors $exp[iS(x_{1}(t))/\hbar]$ and $exp[iS(x_{2}(t))/\hbar]$.
We say that two paths are equivalent, if their corresponding weight factors are identical, i.e. their corresponding action is identical. We can define the distance bewteen two paths by
\begin{equation}
d(x_{1}(t),x_{2}(t)) = |S(x_{1}(t)) - S(x_{2}(t))| ~ .
\end{equation}
It is intuitively plausible that paths with small action $S$ give the dominant contributions to the propagator (otherwise for a path with large action, a little change in the path would lead to strong oscillations and eventually cancel out all those terms). This means also that paths with large fluctuations between neighbor time slices are unfavorable, because they contribute to a large kinetic term in the action. In the case where the potential is of confining type, i.e. $V(x) \to \infty$ when $|x| \to \infty$, then also large fluctuations, where the path deviates much from the classical path are unfavorable, because they give a large potential term in the action. 

The above definition of geometry is purely classical so far: $S$ is the classical action and $x(t)$ is an arbitrary (random) path. Consequently the metric is purely classical. This is changed if we select representative paths 
drawn from a distribution involving the weight factor $exp[iS(x(t))/\hbar]$.
This is exactly what is done when computing the quantum propagator 
(in imaginary time) via Metropolis Monte Carlo. Then 
one finds the action of paths to follow a Gaussian distribution centered around a small value of action. Then the above metric gives some information about 
the distance of quantum mechanical important paths.

A different notion of geometry arises, when we look at the fluctuations due to the kinetic term of the action. A quantitative way to do this is by looking at the Hausdorff dimension $d_{H}$ of an average path $x(t)$.
This can be done by computing the expectation value $<L_{path}>$ of the length of paths averaged over all paths and weighted by $exp[iS(x(t))/\hbar]$.
In a numerical simulation by Monte Carlo one measures
$<L>$ versus the variance $\Delta x$.
In the limit when $\Delta x \to 0$, one obtains a power law of the form
\begin{equation}
<L> \propto \Delta x^{\alpha} ~ ,
\end{equation}
The exponent $\alpha$ determines the Hausdorff dimension ($\alpha = 1 - d_{H}$). More details of numerical experiments can be found in Refs. \cite{kn:Krog95,kn:Krog00}.
Numerical results show $d_{H} =2$, for all local potentials confirming the analytic result by Abbot and Wise \cite{kn:Abbot81}. 
This means the Hausdorff dimension is not sensitive to the interaction.
This is a somehow deceptive result, in the sense, that fractal geometry is not a useful tool for the purpose of a geometrical interpretation of quantum physics. 

However, the above result is valid in flat Riemannian geometry, i.e. in    
absence of any gravitational field. The situation is different, when one considers quantum mechanics in curved space time, i.e. when a quantum particle propagates in the neighborhood of a massive stellar object like a neutron star or a black hole. One expects that in such situation the fractal dimension should differ from $d_{H}=2$.

\section{Quantum mechanics versus neuroscience}

\subsection{Typical scales: Q.M. versus neuroscience}
\label{sec:Typical}
First of all we want to state clearly that neuroscience is generally considered 
as a field where the laws of classical physics apply and the laws of quantum physics are not relevant. In order to get a picture let us look at some typical scales of energy, length and time, being characteristic in quantum physics versus neuroscience. 
In quantum physics the following quantities play a role in setting 
scales in atomic physics: \\
(1) Action: $\hbar = 6.58 ~ 10^{-22} ~ MeV ~ s$. 
This is a fundamental constant of nature. \\
(2) Energy: $E_0$ the ground state energy of an atom. For the hydrogen atom, $E_{0} = -13.6 ~ eV$. \\
(3) Time: $\hbar /E_0$. This sets the time scale in dynamical processes, like a tunneling transition. For the hydrogen ground state 
$\hbar /E_{0} = 0.48 ~ 10^{-16} ~ s$.\\
(4) Length: $\lambda$ the de Broglie wave length of a particle. 
For a thermical electron of kinetic energy $1 ~ eV$, one has $\lambda = 1.95 ~ 10^{-8} ~ cm$. 
Another typical length scale is radius of an atom, e.g., the Bohr radius of the hydrogen atom, which is in the order of 1 Angstrom.

In neuroscience the following scales play a role: \\
(1) Size of central nervous system $1 ~ m$. \\
(1) Size of a neuron $100 ~ \mu m$. \\    
(2) Size of a synapse $1 ~ \mu m $. \\
(3) Typical length of dendrite $1 ~ mm$. \\
(4) Time: duration of action potential $1 ~ ms$. \\
(5) Firing rate $50 ~ Hertz$. \\
(6) Refractory period $1 ~ ms$. \\
(7) Oscillations: $\alpha$, $\beta$, $\gamma$, $\theta$ 5-80 Hertz. \\
Obviously typical length and time scales differ from atomic physics to 
neuroscience.

\subsection{Different concepts of probability: Q.M. versus neuroscience}
\label{sec:Concepts}
The concepts of probability in quantum mechanics and in neuroscience 
are different. In quantum mechanics probability is introduced as 
the interpretation of the absolute square of a transition amplitude or wave 
function. In neuroscience it shows up e.g. in the erratic (noisy) behavior of ions passing membranes via ion channels or in the diffusion of neurotransmitters in synaptic transmission. 
While the wave function obeys the Schr\"odinger equation, the diffusion of 
neurotransmitters obeys the diffusion equation. 
In order to make the distinction clear, let us consider the case of free motion 
(absence of any driving force or potential).
Then the Schr\"odinger equation reads 
\begin{equation} 
- \frac{\hbar}{i} \frac{\partial}{\partial t} \psi(x,t) = - \frac{\hbar^{2}}{2m} \frac{\partial^{2}}{\partial x^{2}} \psi(x,t) .
\end{equation}
On the other hand, the diffusion equation can be written as
\begin{equation}  
\frac{\partial}{\partial t} P(x,t) = D \frac{\partial^{2}}{\partial x^{2}} P(x,t) .
\end{equation}
Here $P(x,t)$ denotes the probability to find the particle at position $x$ 
and time $t$. $D$ detotes a constant, which characterises the diffusion process. Mathematically one observes a great similarity between both equations. 
Actually, the Schr\"odinger equation goes over to the diffusion equation under the transformation
\begin{eqnarray}
t &\longrightarrow& - i t  
\nonumber \\
\frac{\hbar}{2m} &\longrightarrow& D .
\end{eqnarray}
While the second equation is a mere scale transformation, 
the important difference is due to the complex number $i$.
As a consequence the laws of adding probabilities in quantum mechanics and in a diffusion process are totally different. 
This point has been discussed in a most clear and lucid way by Feynman and 
Hibbs \cite{kn:Feynman65}. The mathematical rule of adding probabilities in Q.M. is the following: The basic entity is the wave function or probability amplitude
$\psi$. The wave function at some position $\psi(\vec{x})$ is a complex number.
The probability to find a particle in volume $\Delta V$ around position $\vec{x}$ is given by 
\begin{equation}
\label{eq:Prob1}
\Delta P = |\psi(\vec{x})|^{2} \Delta V ~ .
\end{equation}
In shorthand, the probability is related to the probability amplitude by
\begin{equation}
\label{eq:Prob2}
P = |\psi|^{2} ~ .
\end{equation}
The prescription in quantum mechanics is: One does not add probabilities, but the probability amplitudes.
\begin{eqnarray}
\label{eq:Prob3}
&& \psi_1 ~, \psi_2 \to \psi = \psi_1 + \psi_2 
\nonumber \\
&& P_1 = |\psi_1|^{2}, ~~ P_2 = |\psi_2|^{2} \Longrightarrow
\nonumber \\
&& P = |\psi|^{2} = P_1 + P_2 + 2 Re[\psi_1^{*} \psi_2] 
\neq P_1 + P_2 ~ .
\end{eqnarray}
In diffusion dynamics the rule is  
\begin{equation}
\label{eq:Prob4}
P_1, ~ P_2 \to P = P_1 + P_2 ~ .
\end{equation}
The physical consequence of the diffent laws of adding probabilities can be seen, for instance, in an interference experiment. In quantum mechanics, when a source emits electrons, which pass through a screen with two slits, a detector 
counts intensity (probability) which has a typical shape of maxima alternating 
with minima. The minima are due to destructive interference which is possible due to the term $2 Re[\psi_1^{*} \psi_2]$.    
Contrary to that a particle obeying diffusion dynamics yields an interference 
pattern with a single maximum only (see Ref.\cite{kn:Feynman65}).

One may take a closer look at the law of adding probabilities in Q.M. and ask: Why is it that the wave function has to be complex? After all, some destructive interference might occur already, if $\psi_1$ and $\psi_2$ would be real but of opposite sign. Some indirect evidence is given by the Aharonov-Bohm experiment, where a magnetic solenoid is placed in Young's double slit experiment. Then the wave function reveals topological behavior (i.e. its phase distinguishes if the solenoid is interior or exterior to a closed loop formed by the classical trajectories of two particles going from the source to the detector but traversing different slits). Such behavior is possible only when the wave function is complex. Experimentally, such complex phase factor results in a lateral displacement of the interference pattern.

There is another plausibility argument showing the necessity of the complex unit $i$ to occur in the wave function. Let us consider the time evolution in classical mechanics in a Hamiltonian system with a Hamilton $H$.
It can be expressed by an operator $\exp[\tilde{H} t]$ by
\begin{equation}
q(t) = \exp[\tilde{H} t] ~ q(t=0) ~ ,
\end{equation}
where $\tilde{H}$ denotes a Lie operator, the mapping of which applied to a phase space variable $q$ is given by taking the Poisson bracket of $q$ with the Hamiltonian $H$. 
The previous equation is similar to the evolution 
of the wave function under a (time-independent) Hamiltonian in Q.M.,
\begin{equation}
\psi(t) = \exp[- i H t/\hbar] ~ \psi(t=0) ~ . 
\end{equation}
The difference between both equations is, first of all that $\tilde{H}$ and $H$ operate in different function spaces. But more importantly, the wave function 
guarantees the conservation of probability, which is not the case in classical mechanics (where the Liouville measure is conserved). This means that
$\exp[-i H t/\hbar]$ must be a unitary (or anti unitary operator). But $H$ 
representing the observable energy is Hermitian. The factor $i$ occurs a necessity when relating a Hermitian operator with a unitary operator of exponential form. The same argument applies when considering rotations, being also unitary operators, where the generators are the operators of angular momentum (or spin), which again are Hermitian.

\section{Random behavior in neuroscience}
\label{sec:Randomness}
Neurons are noisy \cite{kn:Ferster96}. First, there is noise in the ion channels. Second, a neuron in the visual cortex, when repeatedly stimulated, never responds in the same way, neither in time nor in amplitude \cite{kn:Tolhurst83}. Another example is synaptic transmission. When a spike arrives from the axon at the presynapse, it does not 
trigger with certainty a signal on the postsynaptic side. The liberation, 
propagation and the process of docking of neurotransmitter molecules is a random process. The neurotransmitter molecules are large molecules (proteins). 

One may ask, if a single neurons acts noisy, how does the brain "know" about the presence of a stimulus? In the brain usually very many neurons respond to a stimulus, hence the brain can filter out a signal from the noise.
This has a mathematical foundation in probability theory and statistics.
The central limit theorem says that when taking the arithmetic mean of a large number $N$ of random variables, the statistical error $\sigma$ of this mean goes to zero like $1/\sqrt{N}$, i.e. behaves more and more deterministically.  
As a consequence, a phenomenological model describing the activity of the membran potential of the neural cell, which involves thousands of ion channels, 
can be well described by a deterministic model, the Hodgekin-Huxley equations.

From this one gets the impression that noise seems to be a nuisance, like in many areas of science. However, there are indications that the brain may also take advantage of the presence of noise. An example is the mechanism of stochastic resonance, which serves as an amplification mechanism of weak signals. This and other examples of the role of noise will be discussed in the following.

\subsection{When noise in neurons plays a constructive role }
\label{sec:NoisyBrain}
In physics, there are many examples, where noise plays a destructive role:
Example: Line broading of a spectral line, due to thermal oscillation of the 
atom which emits the light. Another example: Diffuse backround light being 
emitted from populated areas, which disturbs astronomical observation in the 
night. On the other hand, there are phenomena in nature, which are due to a 
stochastic process or which are coupled to a stochastic process, and where the 
presence of noise plays a constructive role: As a result the signal-to-noise 
ratio is enhanced, or more generally there is creation of order out of disorder.
For example, noise plays an important role in the mechanism of hearing in the ear \cite{kn:Rattay00}. We will discuss the following three mechanisms: \\
(1) Auto-criticality and the sand pile model, \\
(2) Self-organisation off thermal equilibrium: creation of order out of disorder. Examples are the Belousov-Zhabotinski reaction, and Bernard convection. \\
(3) Stochastic resonance and ergodicity breaking in systems with many valley structure.

\subsection{Sand pile model}
\label{sec:SandPile}
The sand pile model is a simple model for a system which keeps some parameter 
at its critical value \cite{kn:SandPile}. It does so without external interference. 
It is a mechanism for self-organized criticality. It explains the $1/f$ noise
observed in transport systems like resistors, the hour glass and luminosity of stars. The $1/f$ behavior reflects a critical state of minimally stable clusters of all length scales. The model has been related to the behavior earthquakes, forest fires, ecology, stock markets, and weather.

Imagine a pile of dry sand. Possibly remote memories from childhood on the 
construction of sand castles tell us that those construction used to decay 
under the influence of sun and wind into a lump of sand with a shape similar 
to a cone. Suppose we have such a cone. Then adding on top a few more grains of sand may trigger an avalanche of sand, such that as outcome a cone shape is restored.
The point is that there is a critical value of steepness, which the sand pile 
tries to maintain. Now the avalanche is a stochastic process, which is needed 
to keep the system at its critical value of steepness. 

Does such mechanism play a role in biology?
First, an ant hill resembles very much to a sandpile, it is conceivable that 
ants (European red would ant) in constructing the ant hill have do deal with critical stability. One may ask: Do ants "know" the critical steepness? 
Second, does such kind of mechanism of self-organized criticality play a role in the working of the brain? This is not known! However, one might speculate about this mechanism because parts of the brain are known to function quite autonomously (breathing, heartbeat).

\subsection{Self-organisation off thermal equilibrium}
\label{sec:SelfOrganis}
Prygogine and collaborators were the first to propose and develop the idea that 
in nature processes off thermodynamical equilibrium occur where locally the entropy decreases leading to states of higher order. In biology this may lead to  forms of life of higher order and complexity.
A fine discussion of self-organisation off thermodynamical equilibrium, 
is given by Nicholis \cite{kn:Nicholis}.
Well known examples are the Belousov-Zhabotinski reaction and Bernard convection. Let us consider the Bernard convection. Anyone who has been heating oil in a frying pan has had the chance to observe this phenomenon: At the beginning when 
the heat has just been turned on, the surface of oil is quiet and flat. After 
some time when the heat is sufficienly strong, one observes that the fluid of 
oil creates honey-comb like structures. They are quite stable.
The physical reason is the difference of temperature between the pan and the 
upper surface of the liquid. It creates a circular motion, which manifests 
itself in an ordered structure. Due to the temperature difference this is an 
off-equilibrium process. It is evidently a random process. Its outcome has a 
non-random deterministic geometrical shape. 
Considering oil, pan and source of heat as a single system, this system manages to create order out of disorder (a signal out of noise) without external interference. 

The Bernard convection occurs also elsewhere in nature, for example in the 
motion of air stream in the upper atmosphere. One may even speculate if the 
formation of galaxies, which is known to have voids and some regular structures \cite{kn:VoidsRegular} may have to do with the Bernard convection mechanism.   
The question is: Is such mechanism pertinent in the dynamics of the brain? 
This is also unknown! However, the brain is known to create spatially and temporally domains of order, where peviously there has been disorder. 
An example is the desease of epilepsy during an attack, where quite regular (orderly) firing patterns are observed in neural activity.

\subsection{Stochastic resonance}
\label{sec:StochRes}
The mechanism of stochastic resonance has been proposed to explain the recurrence of ice ages \cite{kn:Benzi81}. Reviews of the mechanism of stochastic resonance can be found in refs. \cite{kn:Moss95,kn:Wiesenfeld95,kn:Grammaitoni98}.
Longtin \cite{kn:Longtin93} proposed that this mechanism plays a role in excitable systems like neuron models. Douglass et al. \cite{kn:Douglass93}
demonstrated this mechanism at work in mechanoreceptor cells in crayfish. Collins et al. \cite{kn:Collins96} showed that it enhances tactile sensation in man. Also it is present in the sensor neurons of the rat \cite{kn:Nozaki99}, as well as in crickets \cite{kn:Levin96}. 
A recent review on more examples of stochastic resonance is given in 
ref. \cite{kn:Gebeshuber00}.

What is stochastic resonance? As a simple example consider in classical physics 
a particle moving in a double well potential. If the kinetic energy is insufficient to overcome the potential barrier, the particle is confined to stay in the same well all of the time. Now suppose one adds friction, also a periodic force which lowers one potential bottom and raises the other one and after one period goes in the opposite direction. Finally, one adds a random (noisy) force which helps the particle to overcome the barrier. Then an interplay between periodically changing potential, friction and random force creates a "resonant" motion of the particle from one potential bottom to the other. 

How can such mechanism play a constructive role in neuroscience?   
Let us consider a neuron and the creation of an action potential. 
In simple terms the neuron can be viewed as a 2-state system: one state when it fires a spike, and the other state when it is quiet. The mechanical analogue is a system where the particle is in one or the other well. Adding a noisy force may help the mechanical system to more easily go over from one state to the other. Thus the presence of noise in the neuron may change its response to create an action potential. Such noise is present in excitatory synapses.

Another example where noise may play a constructive role, is the associative memory in the brain. The associative memory has been described 
by the Hopfield model. There is an energy surface of configurations (corresponding to an ensemble of neurons, 
each one being in the state of firing or quiescence).
The Hopfield model is mathematically equivalent to a spinglass in condensed matter physics. Spin glasses exhibit a many-valley structure in the 
free energy (broken ergodicity). 
In the associative memory model, the bottom of a valley corresponds to a stored 
pattern, i.e. a piece of stored memory. 
Suppose one wants to go from one pattern of memory to another i.e. from one valley to another. Those who hiked in mountains know that this can cost quite a bit of energy. Such energy may be provided in form of some noise.  
Also when solving such spin glass models numerically, one adds noise to 
facilitate the migration through the whole phase space to search for the global  inimum (algorithm of simulated annealing). May be the brain uses a similar method to switch from one memorized pattern to another pattern. 
 
I summary, noise helps 
to overcome ergodicity breaking (not getting caught in a valley). 
This is likely to be important in the brain. Going from one valley 
to another in the brain means to go from one memorized pattern to 
another memorized pattern. In this picture the role of noise can be seen as a 
motor which helps to populate higher lying energy levels of a spin 
glass system, which corresponds in the associative memory to memorize or access memorized patterns located at higher values of the energy (cost) function.

\section{If neurons would operate deterministically how would the brain look like?}
\label{sec:NonStochasticBrain} 
Let us recall that atomic and molecular physics is based on the concept 
of probability (see sect. \ref{sec:Concepts}). Neuron dynamics shows stochastic behavior in the synaptic transmission and in the opening/closing of ion channels in the membrane. Both processes involve atoms and molecules. 
Thus the probabilistic aspect is inherited from quantum mechanics. 
On the other hand, in neuroscience the diffusion of neurotransmitter molecules 
has been successfully described by diffusion equations, being the limit of Brownian motion. As we pointed out above, the probability concept of quantum mechanics is different from that of Brownian motion.

\begin{figure}[tph]
\vspace{9pt}
\begin{center}
\includegraphics[scale=0.6,angle=0]{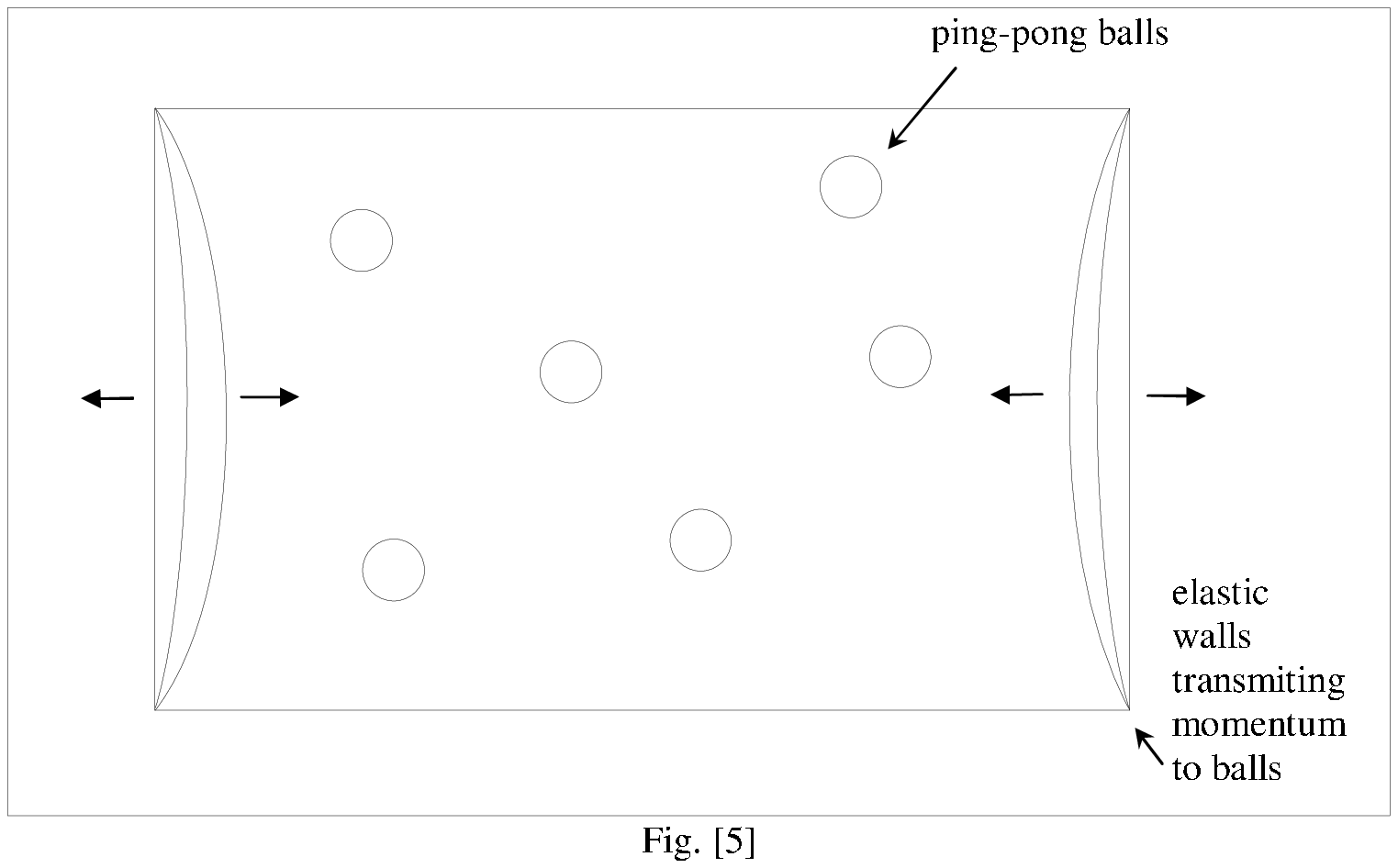}
\end{center}
\caption{
Schema of Brownian motion in macroscopic physics: 
Ping-pong balls in a closed box. 
}
\label{fig:MacroBrown}
\end{figure}

We should note that neurotransmitter molecules are quite large molecules (proteins). We ask: Is Brownian motion a property of microscopic particles, obeying or can macroscopic particles also carry out Brownian motion? We suppose 
the answer is yes! As an example consider ping-pong balls in a large container with elastic walls transmitting momentum to the balls (see Fig.[\ref{fig:MacroBrown}]). 
The mean free path should be much larger than the size of the balls. 
This is a question of density. Brownian motion requires a large number of 
particles to enter in collisions. In the synapses, the collisions occur 
between the neurotransmitter molecules and the liquid in the synaptic cleft. 

So we come back to the question: If a neuron would not operate 
stochastically - described by Brownian motion - how could it operate? Could it work at all? First of all this means that a synapse would be physiologically
different. One can speculate about chemically different neurotransmitter 
molecules, which would be of different size, leading to different 
collision rates and hence different diffusion properties. 
Or it would be conceivable that the noise would be frequency-dependent:
Fast diffusion might correspond to a noise with small sigma (deterministic limit). Or one might consider the fluid in the synaptic cleft to be chemically different, e.g. like a colloid, where neurotransmitter molecules might get 
stuck. All this would lead to a different synaptic transmission behavior.
Ultimately, is it possible that a biological synapse could work deterministically? 
I think the answer is no! The argument is of the same kind as for the 
quantum mechanical decay of a radio-nucleus: It would be too complicated!

\subsection{Dynamic regimes where the brain operates deterministically}
\label{sec:DeterministicBrain}
\begin{figure}[tph]
\vspace{9pt}
\begin{center}
\includegraphics[scale=0.6,angle=0]{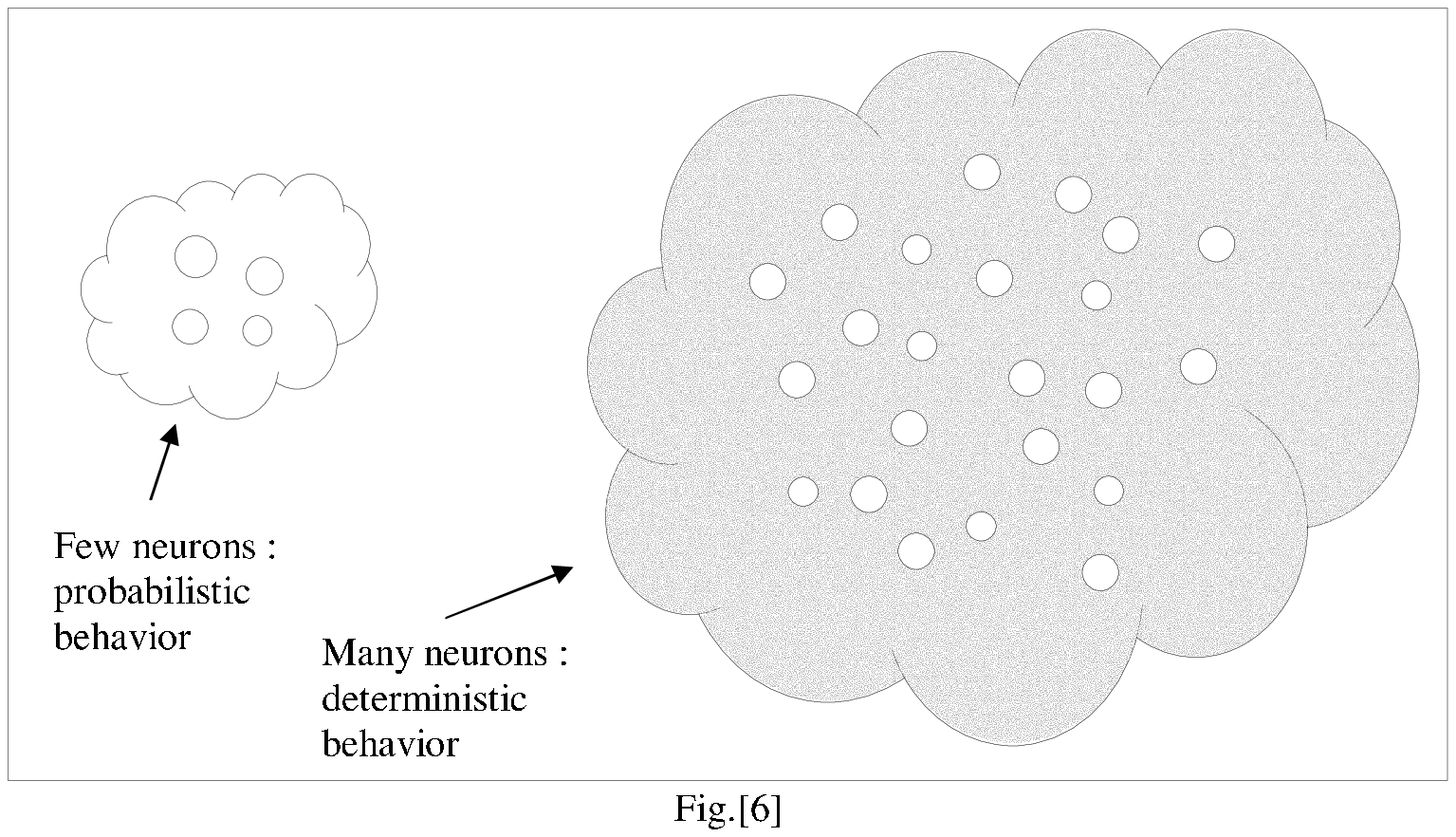}
\end{center}
\caption{
Activation of few neurons gives probabilistic response, 
while coherent activation of many neurons gives deterministic response. 
}
\label{fig:NeurResp}
\end{figure}

Let us consider as example the motoric system of the brain. When 
a fast response to a stimulation is required (an animal flees after an attack from a predator), then many cells fire together. A big muscle undergoes contraction. This requires a coherent effect of many neurons. The ensemble of motor neurons responds in a deterministic way.
This situation is similar to the transition in physics from a few-body system
to many body system. For a many-body system (at thermodynamical equilibrium) 
one can use statistical mechanics. Macroscopic observables behave classically
(Central Limit Theorem). 
This is realized also in the brain (see Fig.[\ref{fig:NeurResp}]). 
How can we tell in which mode the brain is working - probabilistic or deterministic? 

A schematic possible scenarios is the following 
There are many small areas with a few neurons. If the small areas interact 
only weakly then this part of the brain likely works in 
a probabilistic mode. If there is 
one large area where the neurons interact strongly then this part 
of the brain operates coherently (at least for some time). 
Then the brain is in a deterministic mode.

The question is then: What exactly is a strong or weak interaction?
Such questions have been addressed in neural network models. Noest \cite{kn:Noest89} has shown that neural networks can form domains from restricted range interactions.
Recently an interesting answer has been given by the proposal of small world networks and scale-free networks (see sect.~\ref{sec:SmallWorld}).
It has been shown that the small world architecture of neural nets - characterized by strong local clustering and a few short links to distant neural nodes - is capable to yield a fast coherent response \cite{kn:Fernandez00}. 
The issue of the architecture of the brain raises further questions about long range order, phase transitions, order parameters etc. in the brain. Not much is known about this.

\section{Learning}
\label{sec:Learning}
Here we want to discuss the process of learning, retrieval of learned information and their neurophysiological basis using chance and probability. 

\subsection{Cellular basis of learning and formation of memory}
Learning and memorizing are mental abilities found in humans and mammals, but also in animals of simpler organisation having only a number of neurons in the order of $10$ to $10^{5}$. An example is the sea-snail (Aplysia californica). 
This animal shows a simple forms of learning: habituation and sensitisation
\cite{kn:KandelSpek}. Habituation is observed, when the animal is exposed to some stimulation of its breathing organs. It responds by a retraction-reflex
of its gills. When repeating this simulation 10 to 15 times, the retraction-reflex does no longer show up. Even after one hour this reflex occurs only in a much weaker form. The animal has undergone some habituation.
Sensitisation is the opposite effect, where some response of the animal becomes stronger. The molecular mechanism of short term habituation and also sensitisation has been localized to occur in the synaptic transmission.
It is based on a change in the current of $Ca^{++}$ ions passing through ion channels in the membrane of the synaptic boutons of sensory neurons. The concentration of $Ca$ ions in the boutons controls how many synaptic 
vesicles will deliberate their neurotransmitter molecules into the synaptic cleft (space between pre-synaptic and post-synaptic neuron), after an electric action potential has arrived 
at the presynaptic bouton. In the process of habituation, the number of free $Ca$ ions decreases in the boutons, resulting in reduced flow of neuro-transmitter molecules. Eventually the reflex vanishes completely. On the contrary, in the process of sensitisation, the flow of free $Ca$ ions into the synaptic bouton is increased, resulting in an amplified flow of neurotransmitter molecules. The molecular basis is more complicated, requiring 
the interaction with an interneuron (and a chain reaction involving Serotonin,   Adenylat-Cyclase, cyclic Adenosinmonophosphate, Protein-Kinase).

A higher form of learning and memorisation is found in mammals and humans.
It is based on a change (plasticity) in the synaptic transmission. One makes a distinction between short term potentiation (STP) and long term potentiation (LTP). LTP is a long-lasting increase in the amplitude of the synaptic response following brief, high-frequency activity of a synapse. LTP was first described in the hyppocampus by Bliss and Lomo \cite{kn:Bliss}.  
Let us consider as example of LTP the Schaeffer collateral CA1 pyramidel cell in the hyppocampus \cite{kn:Byrne}. A change occurs in the synapse on the post-synaptic side. The origin is an interaction between ionotropic glutamate receptors, AMPA (2-amino-3-propanonic acid) and NMDA (N-methyl-D-aspartate). First the AMPA and NDMA ion channels are closed. When the post-synaptic membrane becomes depolarized, the neurotransmitter interacts with AMPA and NMDA ion channels. Then the AMPA channels open to create an excitatory post-synaptic potential. But the NMDA channel is blocked by $Mg^{2+}$ ions. Only when the post-synaptic membrane is very strongly depolarized, the $Mg^{2+}$ ions are 
liberated and then NMDA channels open. The NMDA channel allows  $Na^{+}$,   $K^{+}$ and $Ca^{2+}$ ions to pass. The $Ca^{2+}$ ions induce a number of biochemical reactions increasing the efficacy of the synapse. 
Also there are hints that this LTP process is accompagnied by a release of nitric oxide (NO), which plays the role of a signal being sent back to the pre-synaptic terminal inducing additional neurotransmitter release \cite{kn:NO}.
The NO molecule was found to enhance transmitter release only if it arrives in coincidence with activity in the pre-synaptic neuron. 

Now we ask: Where in those processes of learning does chance and probability play a role? 
In the case of habituation and sensitisation, the process involves the diffusion of $Ca$ ions through the membrane ion channels. This is a stochastic process. Secondly, controlled by the increase or decrease of $Ca$ ions, there is an increased or decreased number of neurotransmitter molecules, diffusing through the synaptic cleft. Again this is a stochastic process.   
Similarly, in the case of LTP, there are two occasions, where
randomness plays a role. First, it is the diffusion of neurotransmitter and the docking at $AMPA$ and $NMDA$ receptors, which is a stochastic process. Secondly, also the retrograde signal of the $NO$ molecule proceeds via some diffusion process, again subject to the laws of chance.

\subsection{How and where is information stored in the memory?}
We know that learning in humans and most likely also in animals can be distinguished as explicit and implicit learning.
Explicit learning means to remember and recognize people or places. It involves the temporal lobe, the hippocampus (for short term memory) and the  
cortex (for long term memory). Implicit learning means perceptual and motor learning without conscious awareness. It involves the cerebellum and amygdala 
(see Refs.\cite{kn:Kandel,kn:Squire}). 
The process of forming the memory goes through different stages. Recent memory is easily disrupted until the information is converted into a long-term memory. 
Then it is relatively stable. However, when time goes on, the stored informationm and the capacity to retrieve information gradually diminishes. The memory process undergoes a continuous change with time. For different types of learning one knows that the memory is not localized in just one particular place in the brain. For example, any of three visual pathways can sustain conditioning of heart rate response in pigeons.

It is known that the mechanism of learning on the cellular level 
involves change in the synapses.
Implicit learning leads to changes in the effectiveness of the synaptic transmission. Establishing long-term memory requires the synthesis of new proteins and the growth of new synaptic connections. The storage of explicit memory in mammals uses long-term potentiation (LTP) in the hippocampus.
Comparing the storage of memory in an electronic computer with that of 
the brain, one finds a big difference. In a computer information is stored 
locally with precise addresses.
In the brain information is stored non-locally, it is distributed over a large number of synapses and eventually even several parts of the brain.

\subsection{Learning in neural networks}
Learning in neural networks can be done in basically two ways: supervised learning and unsupervised learning. 
The training of neural networks is most efficient, when the learning is supervised. There are many network models using all kinds of variants of supervised learning rules, like feed-forward networks (perceptrons), recurrent networks allowing for connections and information flow in forward and backward direction, or Boltzmann machines (for an overview see \cite{kn:Hertz91}).
However, in neurobiology, a learning process without supervision is more realistic. There is no teacher. This requires some self-organisation of neurons and connections. An learning rule realized in neurons has been proposed in 
1949 by Donald Hebb \cite{kn:Hebb}. He formulated what is known today as Hebb's learning rule: "When an axon of a cell A is near enough to excite cell B or repeatedly or persistently takes part in firing it, some growth process or metabolic change takes place in one or both cells such that A's efficiency, as one of the cells firing B, is increased". Unsupervised Hebbian learning networks have been widely applied to model the visual cortex. 
The formation of orientational columns in the visual field in young cats has been studied by von der Malsburg \cite{kn:Malsburg73} and others.
Linsker \cite{kn:Linsker88} proposed a model of self-organisation of the visual system which does not require structured input, i.e. he used random noise as input of the first layer. It is a muli-layer (modified) Hebbian learning network. As a result he observed in layer three the formation of center-surround cells (maximal response to bright spot in center of recptive field), other cells  showed a "Mexican hat" covariance function (nearby units were positively correlated, while distant units had a negative correlation). In other layers he observed synaptic weights deviating from circular symmetry, although the system 
had a symmetric architecture. It is remarkable that all those features emerged from random noisy input.

\subsection{Neural network model of associative memory}
The associative memory problem is considerd as an example where neural network models have given some insight into neuroscience. The Hopfield model \cite{kn:Hopfield} is considered as a kind of standard model to study this problem. The Hopfield model is given by a Hamilton function of configurations of patterns (composed of bits) plus an up-date rule 
for the evolution of the system. The Hamilton function is very similar to a Hamilton function used in condensed matter physics to describe ferro (or anti-ferro) magnetism. Because the working of memory in a mammalian brain involves a very large number of neurons, one has to consider any associative memory model as a many-body problem. In physics, one often encounters many-body systems. E.g., magnetism, conduction of electrons in metals, or supraconductivity are many-body problems from condensed matter physics.
Thus physicists have developped techniques to mathematically handle and eventually solve (at least approximately) such problems.     
One of those methods is to make use of statistical mechanics. 
In this case one assumes that the system is in a state of statistical equilibrium. This is a very strong assumption, which tells us a rule for the probability that the system can be found in state (configuration) for any given energy. It is the so-called Boltzmann-Gibbs distribution law.
From the point of view of experimental neurobiology, this assumption is a 
gross oversimplification, if not to say, it is simply wrong. Nevertheless, it has turned out to be very useful for the purpose to solve the Hopfield associative memory model. 
This has been achieved by Amit, Gutfreund and Sompolinsky \cite{kn:Sompo,kn:Amit}.
The Hopfield model is mathematically equivalent to a model which describes a spin glass, the Sherrington-Kirkpatrick model. Its Hamiltonian 
\begin{equation}
H = - \sum_{i<j} J_{ij} \sigma_{i} \sigma_{j} 
\end{equation}
is similar to the Ising model of spins $\sigma_{i}$. However, 
the coupling between spins, $J_{ij}$ are considered as random quenched variables, independent for any pair $i,j$ and obeying a Gaussian distribution,
\begin{equation}
P[J_{ij}] = \prod_{i<j} \sqrt{N/2\pi} \exp[- J_{ij}^{2} N/2 ] ~ .
\end{equation}
Now randomness enters here two ways. First, the Hamiltonian has random coupling, which is a characteristic property of spin glasses, creating disorder.  
Second, the free energy function 
\begin{equation}
F = - T \log Z ~ ,
\end{equation}
with $Z$ being the partition function, is a function with a multiple valley structure. The local minima of this function can be interpreted as stored patterns of memory. The problem is to find those local minima. 
Now noise can be used as a helpful tool to solve this problem. A solution algorithm is simulated annealing, a Monte Carlo method, which works by starting from a high temperature, then gradually reducing the temperature allowing to arrive at the absolute minimum of the free energy. 
The idea of this algorithm is quite similar to the mechanism of stochastic resonance.

\subsection{When lost in a foreign city how to find the railway station?}
Suppose a tourist visits a foreign city. He has lost his map. He does not speak the local language. He can not ask anyone for help.
He needs to find the railway station to catch the train. 
He needs to carry out a search and he needs a strategy.
There are several strategies of searching. For example, he could mark down all streets he encounters, walk them along to the end. If he is unsuccessful, he would start with another street, which starts next to his present position.
Alternatively, he could first walk down all streets going in south-north direction then all streets going in west-east direction. Or he could go in a kind of circles around a point, which he considers as a center (e.g the most crowded place he had encountered).
He also could make a random walk, throwing a die at each corner, 
to choose the new direction.
Which strategy would be most successful?
Learning means to (a) try out different searching strategies, (b) memorize them, and (c) evaluate them by ranking their success.
The next time being confronted with a similar problem, like e.g., searching a source of drinking water when lost in wilderness, he would recall his learned search strategy and in particular which were successful and which were not and 
act consequently. 

\begin{figure}[tph]
\vspace{9pt}
\begin{center}
\includegraphics[scale=0.6,angle=0]{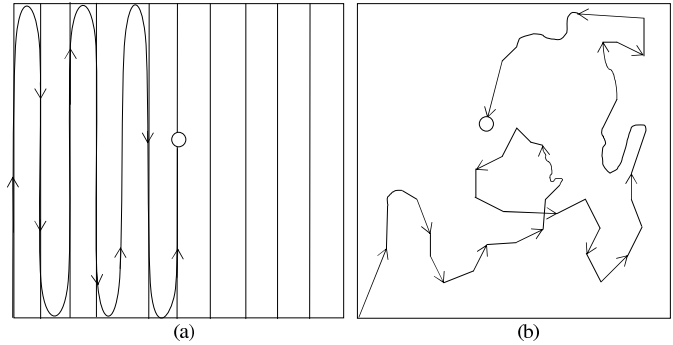}
\end{center}
\caption{
Possible paths in a search: (a) regular path versus (b) random paths 
}
\label{fig:Search}
\end{figure}

Possible search strategies are: (i) Search without guidance. 
One can do a random walk and search or search in a walk following a regular pattern. The success will depend on parameters like step sizes $\Delta L_{step}$, the number of targets $N_{target}$, the distribution of targets.
Also it will depend if one uses a finite target volume $\Delta L_{target}$ (i.e., whenever the searcher has approached the target within a distance $d \leq \Delta L_{target}$, it means he has found the target) and also on the dimension of space (see Fig.[\ref{fig:Search}]). 
(ii) Search with guidance. For instance, a bird of prey recognizes its target from some distance (which may be quite large). Insects are guided by the fragrance of flowers or insects by pheromones. 

For instance if the target can be expressed as a local minimum of some potential or cost function, this represents a standard problem in mathematical optimisation theory.
There are standard algorithms, like the steepest descent method, the conjugate gradient method, the method of simulated annealing or variants of genetic algorithms. In any case, searches with randomness/noise involved will in general be much more efficient (see sect.~\ref{sec:Algorithms}).

\section{Small World and Scale Free Networks}
\label{sec:SmallWorld}
The working of the brain it is not only determined by the number of neurons, but also - among other factors - by the "wiring" i.e. the connections of neurons. Recently, a new type of network architecture has been focused on - the so called "small world" networks \cite{kn:Watts98} and also a variant the "scale-free" networks \cite{kn:Barabasi99}. The small world networks are characterized by 
high local clustering. This means if node A is linked to node B and A is also linked to node C, then there is a high probability that B is linked to C.
An equally important property is that there are short links to distant nodes. 
One can visualize such network architecture by thinking of nodes on a rectangular grid. Each node on a grid point is linked to his next-neighbor node on the grid (clustering). In addition there are a few links connecting pairs of nodes quite distant on the grid (short connections). The network architecture is somewhere between regular and random. The scale-free networks are characterized by many nodes having few links, some nodes having more links and few nodes having many links, the distribution being given by a power law.
Those networks have been shown to explain the Milgram letter experiment.
Such network architectures have been identified to occur in the internet and world wide web, in the distribution of powerlines in the US, in the biology of cellular and metabolic networks (of the nematode worm C. elegans, thoroughly studied in genetics). As an example from neuroscience, also the neural net of C. elegans follows such network architecture.  

What has all this to do with probability and neuroscience? The clue is that this network has some random connections which apparently makes it very efficient. In other words, a purely regular network or a purely random network would be less efficient. This shows up in the fact that small world networks minimize the search time of addresses of nodes. Apparently, at least for the nervous system of C. elegans, small-world is the optimal architecture.  
Also Hodgekin-Huxley neurons have been investigated in computer simulations. 
It turned out that such neurons work optimal i.e. produce coherent oscillations 
and a fast system response, when they are linked in a small world topology \cite{kn:Fernandez00}. This architecture has been explored also in computer simulations of learning. Finally, the small world network was found to be very efficient for the associative memory \cite{kn:Bohland01}.

\section{Algorithms}
\label{sec:Algorithms}
In mathematics it occurs often that deterministic problems can be solved efficiently via probabilistic/stochastic methods.
As a first example consider integrals. In particular, consider integrals where the boundary of the integration domain is not a smooth surface, but has an irregular (crinkly) shape. In such cases a Monte Carlo computation using random nodes is often more efficient than using integration with nodes adapted to the geometry.
Second, and more importantly, the stochastic computation of integrals is very effective (and in many cases the only way) to compute integrals in high dimensions. By a rule of thumb, when the dimension of the integration domain is larger than $d=10$, then a Monte Carlo evaluation of the integral is more effecient than an evaluation using fixed node rules \cite{kn:James80}.
In physics it is a standard method to compute path integrals (where the integration domain has a dimension $d$ in the order of a few thousand) via the Monte Carlo with importance sampling. A widely used algoritm is the Metropolis algorithm \cite{kn:Metropolis53}.

As a second example consider search an optimisation problems, like finding the 
ground state of a Hamiltonian in a high-dimensional configuration space, or of a highly disordered system like a spin-glass, or finding the shortest path in the travelling salesman problem. For such problems, variants of two methods are widely used and quite successful. One is the method of simulated annealing, a variant of the Monte Carlo Metropolis method. The other is the use of genetic algorithms. Both explore and search the configuration space using random numbers. For a comparison between both methods see \cite{kn:Thompson00}.

As a third example consider deterministic differential equations, which can be expressed in terms of and solved by a stochastic process. That means, one can obtain the solution of a deterministic differential equation from a solution of a stochastic process. As a simple example think 
of the diffusion or heat equation and its solution in terms of 
a Monte Carlo simulation (Gaussian process, Brownian motion).
It has been shown by Courant, Friedrich and Levy \cite{kn:Courant28}
that the solution of certain differential equations is equivalent to a random walk. As example consider the diffusion equation in $d$ dimensions,
\begin{equation}
\Delta P(x,t;x_{0},t_{0}) - \frac{1}{D} \frac{\partial}{\partial t} P(x,t,x_{0},t_{0}) = 0 ,
\end{equation}
imposing the initial condition
\begin{equation}
\lim_{t \to t_{0}} P(x,t; x_{0},t_{0}) = \delta(x - x_{0}) .
\end{equation}
$D$ is the diffusion coefficient.
It is well known (for a proof see Ref.\cite{kn:Itzykson})
that the solution $P$ can be obtained from a Gaussian random walk on a spatial regular lattice (lattice spacing $a_{s}$), where the time progesses also in discrete units ($a_{t}$). The solution of the differential equation is obtained
from the solution of the random walk in the limit 
\begin{equation} 
a_{s} \to 0, ~~~ a_{t} \to 0, ~~~ \frac{a_{s}^{2}}{a_{t}} = 2 d ~ .
\end{equation}

What have stochastic algorithms to do with neuroscience? The neuron performs tasks similar to a processor in a computer. 
As an example, the brain is very good at pattern recognition - a baby very early recognizes its mother. Pattern recognition is a cognitive task which can be 
formulated as a neural computation algorithm.
Another example is the creation of an action potential. The neuron, according to the integrate-and-fire model works 
like a cash register with a summing device. It sums the action potentials incoming via dendrites from neighbor neurons. 
As a last example, recall that the synaptic neuro transmitter diffusion process 
is mathematically equivalent to the solution of some differential equation. 

Given the fact that the neuron executes mathematical algorithms, like 
integration or solving differential equations,
it is quite likely an advantage for the neuron to use algorithms based on stochastic methods (like on a computer with electronic processors). 
Note, however, that the neuron is not a digital computer, but rather an analogue one. So the algorithm is directly linked to and manifested in the architecture of a neuron. The conclusion, however, is the same: The stochastic behavior of a neuron is an algorithmic advantage in performing its tasks.
This advantage may show up in faster execution of tasks or in the ability to perform several operations in parallel.

\section{Conclusion}
\label{sec:Conclusion}
In Quantum mechanics, the concept of probability is fundamental. It is related to the existence of stable atoms, hence macromolecules and organic life.
In neuroscience there is a vast number of viewpoints, where noise and stochastic behavior is beneficial. It is tempting to speculate how a brain relying only on a deterministic mode would look like. Quite likely it would not work!

\vskip 1cm

\noindent {\bf Acknowledgments} \\
H.K. has been supported by NSERC Canada.
H.K. is grateful for the kind hospitality at the SALK Institute, where part of this work has been done. H.K. is grateful for discussions with 
Prof. Francis Crick, SALK Institute, 
Prof. Terrence Sejnowski, SALK Institute, 
and Prof. Christof Koch, CalTech.

\end{document}